\documentclass[aps,prl,reprint,superscriptaddress]{revtex4-1}

\usepackage{textcomp}
\usepackage{upgreek}
\usepackage{array}

\usepackage{graphicx}
\usepackage{float}
\usepackage[none]{hyphenat}
\usepackage{amsmath}
\usepackage{epstopdf}
\DeclareMathOperator{\tr}{tr}

\begin{document}


\title{$h/e$ superconducting quantum interference through trivial edge states in InAs}


\author{Folkert K. de Vries}
\author{Tom Timmerman}
\affiliation{QuTech and Kavli Institute of Nanoscience, Delft University of Technology, 2600 GA Delft, The Netherlands}

\author{Viacheslav P. Ostroukh}
\affiliation{Instituut-Lorentz, Universiteit Leiden, P.O. Box 9506, 2300 RA Leiden, The Netherlands}

\author{Jasper van Veen}
\author{Arjan J. A. Beukman}
\author{Fanming Qu}
\author{Michael Wimmer}
\affiliation{QuTech and Kavli Institute of Nanoscience, Delft University of Technology, 2600 GA Delft, The Netherlands}

\author{Binh-Minh Nguyen}
\author{Andrey A. Kiselev}
\author{Wei Yi}
\author{Marko Sokolich}
\affiliation{HRL Laboratories, 3011 Malibu Canyon Road, Malibu, California 90265, USA}

\author{Michael J. Manfra}
\affiliation{Department of Physics and Astronomy and Station Q Purdue, Purdue University, West Lafayette, Indiana 47907, USA}

\author{Charles M. Marcus}
\affiliation{Center for Quantum Devices, Niels Bohr Institute, University of Copenhagen, 2100 Copenhagen, Denmark}

\author{Leo P. Kouwenhoven}
\email{l.p.kouwenhoven@tudelft.nl}
\affiliation{QuTech and Kavli Institute of Nanoscience, Delft University of Technology, 2600 GA Delft, The Netherlands}
\affiliation{Microsoft Station Q Delft, 2600 GA Delft, The Netherlands}

\date{\today}
\begin{abstract}
Josephson junctions defined in strong spin orbit semiconductors are highly interesting for the search for topological systems. However, next to topological edge states that emerge in a sufficient magnetic field, trivial edge states can also occur. We study the trivial edge states with superconducting quantum interference measurements on non-topological InAs Josephson junctions. We observe a SQUID pattern, an indication of superconducting edge transport. Also, a remarkable $h/e$ SQUID signal is observed that, as we find, stems from crossed Andreev states.
\end{abstract}


\maketitle

Topological systems are a hot topic in condensed matter physics~\cite{Kane_2010}. This is largely motivated by the existence of states at the interface between two topologically distinct phases, for example helical edge states in a quantum spin Hall insulator (QSHI)~\cite{Kane_2005,Konig_2007}.  Inducing superconductivity in these edge states would form a topological superconductor~\cite{Kane_2010}. Superconducting edge transport has already been found in materials that are predicted to be QSHI~\cite{Hart_2014,Pribiag_2015}. However, edge states can also have a non-topological origin. Trivial edge conduction is found in InAs alongside the chiral edge states in the QH regime~\cite{vanWees_1995} and recently in the proposed QSHI InAs/GaSb as well~\cite{Nguyen_2016,Nichele_2016}. To be able to discriminate between topological and trivial states it is crucial to study transport through trivial edges also and clarify differences and similarities between them. In this work we study the superconducting transport through trivial edge states in non-topological InAs Josephson junctions using superconducting quantum interference (SQI) measurements. We find supercurrent carried by these edge states and an intriguing $h/e$ periodic signal in a superconducting quantum interference device (SQUID) geometry.

\begin{figure}[h]
	\centering
	\includegraphics[width=3.4in]{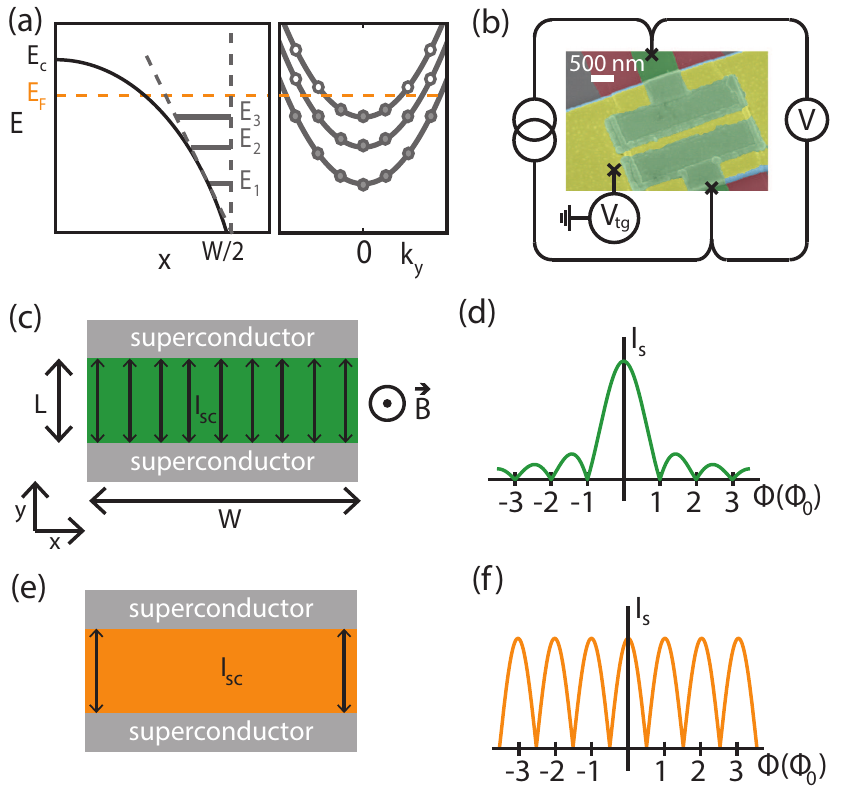}
	\caption{(a) Sketch of the conduction band minimum around the edge of a 2DEG with Fermi level pinning at $W$/2. The band bending leads to a roughly triangular quantum well in the vicinity of the edge, therefore one-dimensional sub bands form of which three are drawn, as an example. The orange dashed line indicates the Fermi level corresponding to the current distribution in (e). (b) False coloured SEM image of the device with dimensions $W$ = 4~\textmu m and $L$ = 500 nm , where the quasi-four terminal measurement setup is added. Red is the mesa, green the NbTiN contacts, blue SiN$_x$ dielectric and yellow the gold top gate. (c) Schematic representation of a Josephson junction of width $W$ and length $L$. A homogeneously distributed supercurrent $I_{sc}$ is running through the whole junction, resulting in (d) a Fraunhofer SQI pattern. (e) If supercurrent only flows along the edges of the sample, (f) a SQUID pattern is observed.
	}
	\label{SQI:fig1}
\end{figure}

Trivial edge states arise when the Fermi level resides in the band gap in the bulk, while being pinned in the conduction band at the surface. Then, band bending leads to electron accumulation at that surface as schematically drawn in~Fig.\,1(a). The Fermi level pinning can have several origins: truncating the Bloch functions in space~\cite{Tamm_1932,Shockley_1939}, a work function difference~\cite{Bardeen_1947}, the built-in electric field in a heterostack~\cite{Furukawa_1993} and the surface termination~\cite{Olsson_1996}. In our 2D InAs Josephson junctions the accumulation surface is located at the edge of the mesa that is defined by wet etching. The quantum well is MBE grown on a GaSb substrate serving as a global bottom gate~\cite{SM}. The superconducting electrodes are made of sputtered NbTiN with a spacing of 500 nm and a width of 4~\textmu m. A SiN$_x$ dielectric separates the top gate from the heterostructure. Electrical quasi-four terminal measurements [see Fig.1(b)] are performed in a dilution refrigerator with an electron temperature of 60 mK unless stated otherwise. 

\begin{figure}[!ht]
	\centering
	\includegraphics[width=3.3in]{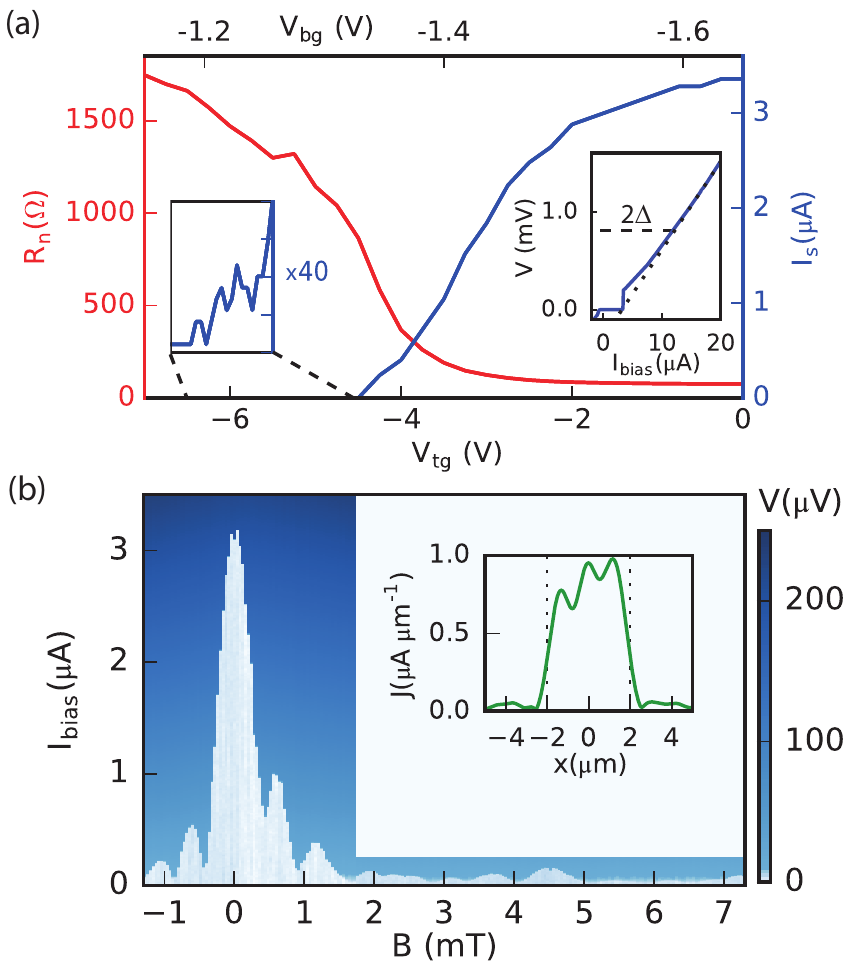}
	\caption{(a) Normal state resistance $R_n$ and switching current $I_s$ at the respective top gate $V_{tg}$ and bottom gate $V_{bg}$ voltages. The left inset depicts a seperate measurement at the indicated gate voltages, where a smaller current bias step size is used for higher resolution. The right inset shows an IV trace at $V_{tg}$ = 0 V and $V_{bg}$ = -1.65 V , where two dashed lines are added for extraction of the induced superconducting gap $\Delta$ and the excess current. (b) The measured voltage as function of the applied current $I_{bias}$ and perpendicular magnetic field $B$ at $V_{tg}$ = 0 V and $V_{bg}$ = -1.65 V. The inset depicts the calculated supercurrent density along the width of the device that is indicated by the dotted lines.}
	\label{SQI:fig2}
\end{figure}

The electron density in the InAs quantum well is altered by using the electrostatic gates, $V_{tg}$ and $V_{bg}$, located above and below the 2DEG. Decreasing the density subsequently increases the normal state resistance $R_n$ and reduces the switching current $I_s$ as shown in Fig.\,2(a). A full resistance map as a function of top and bottom gate is shown in the Supplemental Material~\cite{SM}. The Josephson junction is first characterized at $V_{tg}$ = 0~V and $V_{bg}$ = -1.65~V, where the largest switching current is observed. From the IV trace in Fig.\,2(a) we estimate an induced superconducting gap of 0.4 meV and, using the corrected OBTK model~\cite{Flensberg_1988}, a transmission of $T$ = 0.73. The junction is quasi-ballistic because the mean free path of 2.8~\textmu m (extracted from a Hall bar device on the same wafer~\cite{SM}) is  larger than its length $L$ of 500~nm. The large superconducting gap and high transmission value indicate a high quality InAs Josephson junction.

SQI measurements have successfully been used before to gather information on the supercurrent density profile in Josephson junctions~\cite{Hart_2014,Pribiag_2015,Allen_2016}. This is typically done, using Dynes-Fulton approach~\cite{Dynes_1971}, which connects critical current dependency on magnetic field $I_c(B)$ and zero-field supercurrent density profile $j(x)$ with a Fourier transform. It was originally developed for tunnel junctions, but can also be applied to transparent junctions under several assumptions. Firstly, we should have a sinusoidal current-phase dependency, which is in accordance with the transmission value and temperature in our experiment~\cite{Haberkorn_1978}. Secondly, the Andreev levels, that carry supercurrent in the junction, may only weakly deviate from the longitudinal propagation. Our junction satisfies this constraint since the superconducting coherence length $\zeta = \hbar v_F/\Delta \approx$ 1.3~\textmu m $>L$~\cite{Hui_2014,SM}. If both assumptions hold, we expect Fraunhofer SQI pattern in the case of homogeneous current distribution~(Fig.\,1(c-d)) and SQUID pattern in the case of current flowing along the edges~(Fig.\,1(e-f)). 

A SQI measurement at the largest switching current reveals a Fraunhofer like pattern as shown in~Fig.\,2(b). The central lobe is twice as wide as the side lobes and the amplitude decreases as expected. The slight asymmetry in the amplitudes we contribute to breaking of the mirror symmetry of the sample in the direction along the current~\cite{Rasmussen_2016}. The effective length of the junction [$\lambda=\delta B_{lobe}/(\Phi_0 \cdot W)$] of 1.2~\textmu m is extracted from the periodicity of the SQI pattern. Flux focusing due to the Meissner effect causes it to be larger than the junction length ($\lambda>L$)~\cite{Suominen_2016}. The extracted current density profile, plotted in~Fig.\,2(b), is close to uniform. The supercurrent is thus dominated by bulk transport as expected at these gate voltages. 

The interference pattern in Fig.\,2(b) deviates from the expected pattern after the second lobe. Recently a similar distorted Fraunhofer tail was observed and discussed in graphene \cite{BenShalom_2015}. The perpendicular magnetic field exerts a Lorentz force on the electron and holes suppressing the formation of Andreev bound states. The suppression becomes relevant at a magnetic field scale of $\Delta/eLv_F$, equal to 1 mT in our case, roughly agreeing with the observation. The analysis only holds for the bulk of the junction, since the scattering at the edges reduces the difference in the electron and hole motion in a magnetic field.

\begin{figure}[ht]
	\centering
	\includegraphics[width=3.4in]{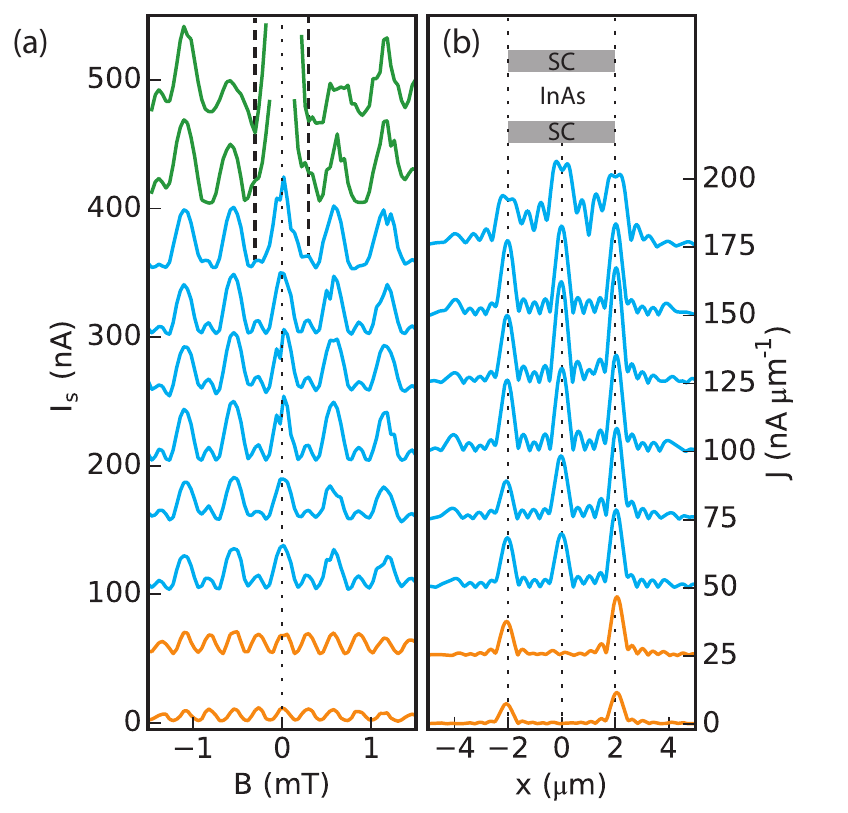}
	\caption{(a) The switching current plotted as function of perpendicular magnetic field and (b) the corresponding current density along the width of the device (see inset), assuming the validity of the Dynes-Fulton approach. The gate values used are from bottom to top: $V_{tg}$ -5.4~V to -3.6~V (0.2~V step) and $V_{bg}$ -1.270~V to -1.396~V (0.014~V step). The green, blue and orange traces are Fraunhofer, even-odd and SQUID patterns, respectively. Since the current is only swept up to 100 nA, the green traces are not suitable for extracting a supercurrent density profile. The traces are offset by 50 nA in (a) and 25 nA/\textmu m in (b). }
	\label{SQI:fig3}
\end{figure}

Next we study the SQI pattern as the Fermi level is decreased by tuning the top gate to more negative values. The upper two (green) traces in~Fig.\,3(a) have a wide central lobe, identifying a Fraunhofer pattern. The effective length is $\lambda$ =1.7~\textmu m, different from before, which we believe is due to different vortex pinning because of the larger magnetic field range of the measurement~\cite{SM}. In the third (first blue) trace we observe that the first nodes turn into peaks, which is highlighted by the dashed lines. This is the transition from a Fraunhofer to a SQUID pattern. Curiously the amplitude and width of the peaks are alternating in the blue traces in~Fig.\,3(a).  The even-odd pattern is composed of an $h/e$ and $h/2e$ periodic signal. An even-odd pattern was observed before in Pribiag  \textit{et al.}~\cite{Pribiag_2015}. In comparison, in this work the amplitude difference in the lobes is much larger and the pattern is visible over a large gate range. The calculated supercurrent density profiles in~Fig.\,3(b) have a central peak that is physically unlikely considering the device geometry. The cause of this intriging interference pattern will be discussed in more detail later. Reducing $V_{tg}$ further we find a clear $h/2e$ periodic SQUID interference pattern in the bottom two (orange) traces. This is a strong indication of edge conduction in our device. Confirmed by the edge transport only in the extracted supercurrent density profiles in~Fig.\,3(b). The transition from bulk to edge transport as a function of gate voltage is measured in several other Josephson junctions~\cite{SM}. Since we observe supercurrent through the trivial edge states of an InAs quantum well, we conclude that a clear demonstration of superconducting edges alone does not prove induced superconductivity in topological edge states.

\begin{figure}[t]
	\centering
	\includegraphics[width=3.4in]{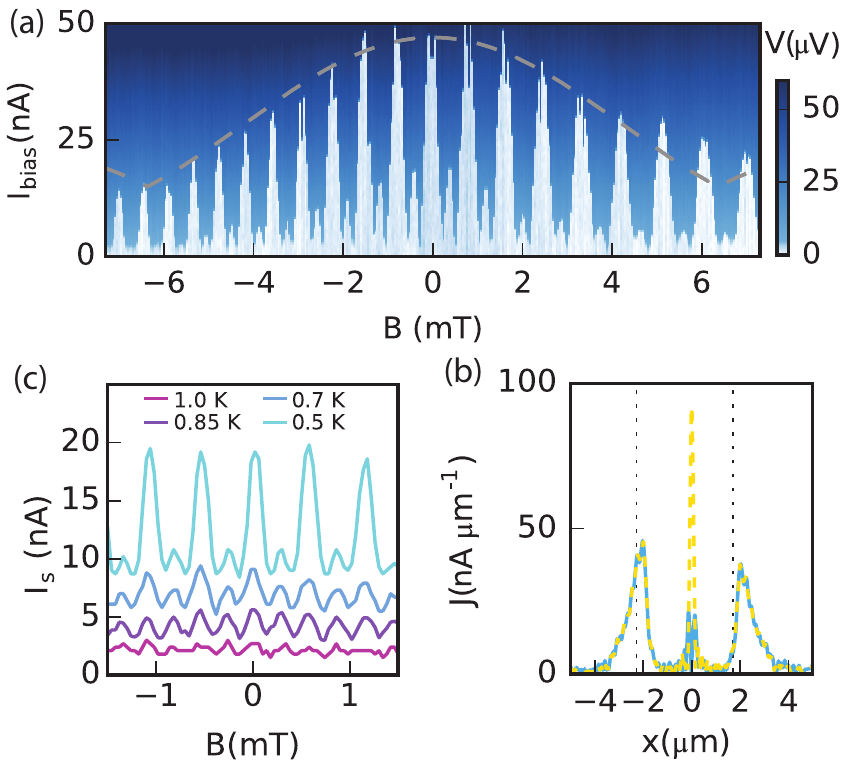}
	\caption{(a) Measured voltage as a function of $I_{bias}$ and magnetic field $B$ at $V_{tg}$ = -5 V and $V_{bg}$ = -1.29 V. (b) Switching current versus the magnetic field for different temperatures at the same gate voltages as (a). The traces are offset by 5 nA for clarity. (c) Current density profile, calculated from the SQI pattern of (a) (see also \cite{SM}). The blue trace uses equation (1), thus correcting the vertical offset in the SQI pattern. The yellow dashed trace is extracted without this correction.
	}
	\label{SQI:fig4}
\end{figure}

We now return to the remarkable $h/e$ SQUID signal to investigate its origin. Figure~4(a) shows a more detailed measurement in this gate regime, the even-odd pattern is observed over more than 25 oscillations. The envelope of the peaks is attributed to the finite width of the edge channels. The effect is suppressed by raising the temperature [see~Fig.\,4(b)], for $T >$ 850 mK a regular $h/2e$ SQUID pattern remains. The origin can not lie in effects that occur beyond a certain critical magnetic field, like $0-\pi$ transitions~\cite{Yokoyama_2014}, edge effects~\cite{Heida_1998, Meier_2016} and a topological state, because we observe the even-odd pattern around zero magnetic field as well. An effect that does not rely on magnetic field and is strongly temperature dependent is crossed Andreev reflection~\cite{Russo_2005}.

The lowest order crossed Andreev reflection (up to electron-hole symmetry) is schematically depicted in~Fig.\,5(a). An electron travels along one edge, whereafter a hole is retroreflected over the other edge. This process alone is independent of the flux through the junction, but still adds to the critical current~\cite{Baxevanis_2015}. Higher order processes that include an electron that encircles the junction completely pick up an $h/e$ phase when a flux quantum threads through the junction, hence the supercurrent becomes $h/e$ periodic~\cite{vanOstaay_2011}. Additionaly, interference processes between crossed Andreev and single edge Andreev states could lead to a $h/e$ contribution~\cite{Recher_2001}. It is important to note that the critical current is $h/e$ periodic in flux trough the sample, but still $2 \pi$ periodic in superconducting phase difference.

Forming crossed Andreev states in the junction is only possible if the particles can flow along the contacts. Electrostatic simulations indeed show a large electron density close to the contacts at gate voltages where the bulk is already depleted~\cite{SM}, because of local screening of the top gate. Nevertheless the needed coherence length for a crossed Andreev reflection is on the order of 10~\textmu m, where the estimated superconducting coherence length (from bulk values) is 1.3~\textmu m. The difference between these values remains an open question.

\begin{figure}[t]
	\centering
	\includegraphics[width=3.4in]{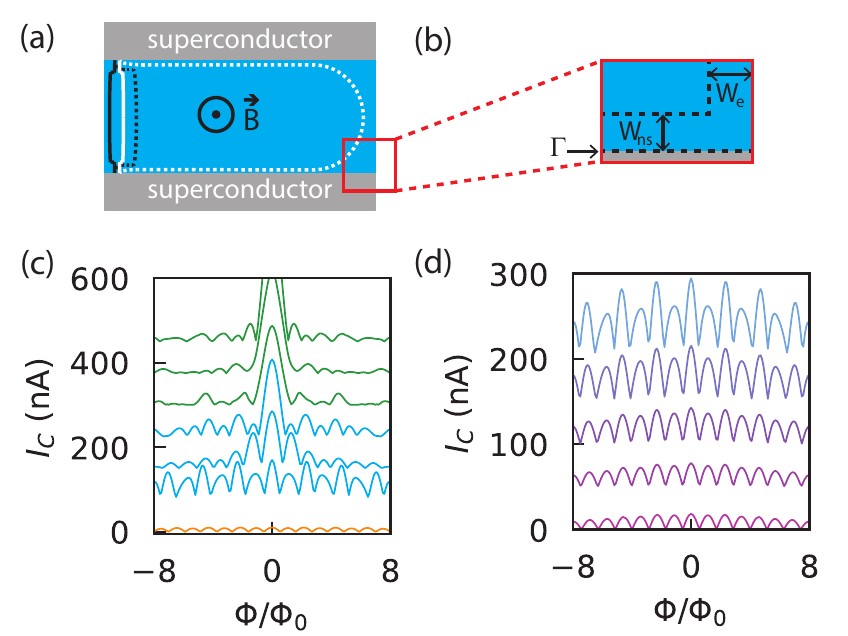}
	\caption{(a) Schematic representation of two crossed Andreev processes. The black and white lines indicate electron and hole trajectories or vice versa. The solid lines represent a single edge Andreev state and the dotted lines a crossed Andreev state. (b) Detailed sketch of one corner of junction in our tight binding mode indicating the widths $W_{ns}$ and $W_e$, and tunnel barrier $\Gamma$. (c) Calculated SQI patterns at overall chemical potential ranging from -0.06~eV to 0.18~eV (0.04~eV step) at 0.46~K and (d) at temperatures 0.4~K, 0.9~K,  1.4~K, 1.9~K, 2.3~K at a chemical potential of -0.2~eV. Traces are offset by 10~nA for clarity. In (c) the color represents the type of interference pattern, green for Fraunhofer, blue for even-odd and orange for SQUID, respectively.
}
	\label{SQI:fig5}
\end{figure}

The phenomenological model proposed by Baxevanis \textit{et al.} considers both single edge and crossed Andreev states \cite{Baxevanis_2015}. In our device we expect the lowest order crossed Andreev states to contribute most because of the short coherence length. Combining their flux insensitive contribution to the critical current and the usual $h/2e$ periodic contribution from single edge Andreev bound states, the model predicts an even-odd or $h/e$ SQUID pattern: 
\begin{equation}
I_c(\Phi)=I_{0}\left|\cos(\pi \Phi/\Phi_0)+f\right|.
\label{SQI:SQUID}
\end{equation}
Where $I_{0}$ the critical field at zero magnetic field and $\Phi$ is the applied flux. Constant $f$ can be arbitrarily large, it dependes on the ratio $\Gamma$ between the probability to Andreev reflect on a node versus the probability to scatter to another edge and is exponentially suppressed by the width of the sample:
\begin{equation}
  f \sim \Gamma^{-1} \frac{k_B T}{\Delta} e^{-2\pi (k_B T / \Delta) (W / \zeta)}.
\end{equation}
The predicted pattern is thus the absolute value of a vertically offsetted cosine function. That is exactly the pattern we measured in~Fig.\,3(a) and 4(a) as both the amplitude and width of the lobes alternate (see also~\cite{SM}). From the data we estimate $f$ = 0.3 and, using the other known parameters, find $\Gamma \sim 10^{-1}$. Taking the Fourier transform in the Dynes-Fulton analysis, offset $f$ leads to a non-physical current density around zero, like we observe in the current density profiles in~Fig.\,3(b) and the yellow dashed line in~Fig.\,4(c). Moreover, the Dynes-Fulton approach is not valid here since crossed Andreev reflection does not meet the second assumption of having straight trajectories only. We can compensate the crossed Andreev contribution by subtracting the constant offset of $f \cdot I_0$=11~nA. This results in a current distribution with mainly current along the edges, as plotted in the blue trace of~Fig.\,4(c). We did not take into account that $I_0$ is actually not constant due to the Fraunhofer envelope of the SQI pattern, so the current density in the center of the junction is not entirely eliminated.

Even though the SQI pattern from the phenomenological model is in qualitative agreement with our data, we also present a tight binding model of system in order to connect it directly to experimentally accessible parameters. In the microscopic model we include the superconducting gap as measured, the width of the paths along the contacts $W_{ns}$ of 20~nm [extracted from the Fraunhofer envelope in~Fig.\,4(a)], and Fermi level pinning on the edges leading to edge current in the region~$W_e$. It is crucial to also take into account a tunnel barrier~$\Gamma$ at the contacts that has a magnitude consistent with the measured transmission value. This barrier enhances normal reflection and therefore elongates the length electrons and holes travel before Andreev reflecting~\cite{SM}. Incorporating these experimental values we find an $h/e$ SQUID pattern. Emulating the experimental gating effect by changing the overall chemical potential results in a crossover from even-odd to Fraunhofer [Fig.\,5(c)], consistent with the measurement in~Fig.\,3. As a check, $W_{ns}$ is reduced in steps to zero, which results in a SQUID pattern~\cite{SM}. Additionally, in~Fig.\,5(d) we observe that increasing the temperature indeed smears out the even-odd pattern and leaves us with a regular SQUID pattern, similar to the experimental data in~Fig.\,4(b). Summarizing, both the phenomenological model and the microscopic model support our hypothesis of the $h/e$ SQUID originating from crossed Andreev states.

We have experimentally shown that trivial edge states can support highly coherent superconducting transport that also becomes visible in an $h/e$ periodic SQI pattern. Both effects have been considered as possible signatures for inducing superconductivity in topological edge states before~\cite{Hart_2014,Pribiag_2015}. Therefore we conclude that superconducting edge transport and an $h/e$ SQUID pattern only, cannot distinguish between topological and trivial edge states, nor can it be considered a definite proof for a topological phase.\\

\begin{acknowledgments}
We thank Daniel Loss for fruitful discussions and Carlo Beenakker and Michiel de Moor for useful comments on the manuscript. This work has been supported by funding from the Netherlands Organisation for Scientific Research (NWO), Microsoft Corporation Station Q, the Danish National Research Foundation and an ERC Synergy Grant.
\end{acknowledgments}

\bibliography{h_e_SQI_deVries_bib}

\onecolumngrid
\linespread{1.5}

\setcounter{figure}{0}

\makeatletter 
\renewcommand{\thefigure}{\normalsize\textbf{S\arabic{figure}}}
\renewcommand{\thetable}{\normalsize \textbf{S\arabic{table}}}
\makeatother

\renewcommand\figurename{\normalsize \textbf{Figure}}
\renewcommand\tablename{\normalsize\textbf{Table}}

\newpage

\begin{center}
\qquad \\[0.1cm] \LARGE{ \huge Supplementary Information \\[5mm] \LARGE $h/e$ superconducting quantum interference through trivial edge states in InAs }\\[1cm]

\linespread{1.5}
\large \noindent Folkert K. de Vries, Tom Timmerman, Viacheslav P. Ostroukh, Jasper van Veen,  Arjan J. A. Beukman, Fanming Qu, Michael Wimmer, Binh-Minh Nguyen, Andrey A. Kiselev,  Wei Yi, Marko Sokolich, Michael J. Manfra, Charles M. Marcus, and Leo P. Kouwenhoven\\[2cm]
\end{center}

\normalsize
\clearpage

\section{Heterostructure stack}
\begin{figure}[h]
	\centering
	\includegraphics[width=2 in]{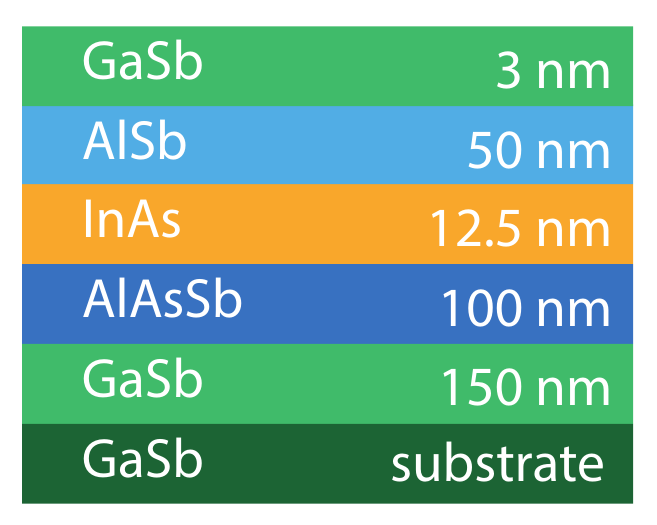}
	\caption{The heterostructure is MBE grown and consists of the following layers: a doped GaSb substrate; a 150 nm GaSb bufffer; the quantum well with an AlAsSb barrier of 100 nm, 12.5 nm InAs and a 50 nm AlSb top barrier; and a 3 nm GaSb capping layer. The latter is used to prevent oxidation of the AlSb. This stack, especially the capping layer, leads to unintentional doping in the InAs quantum well \cite{Furukawa_1993_SM}.
	}
	\label{SQI:figS1}
\end{figure}

\section{Resistance map}
\begin{figure}[!ht] 
	\centering
	\includegraphics[width=3.4 in]{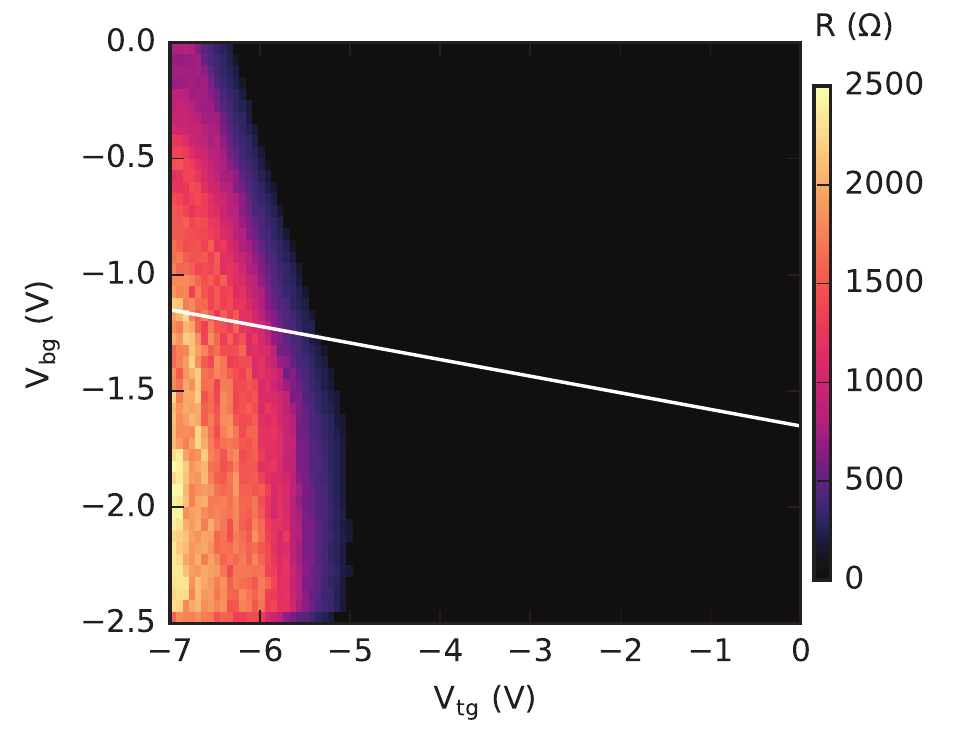}
	\caption{Two terminal resistance of the Josephson junction is plotted as a function of top gate and bottom gate voltage ($V_{tg}$ and $V_{bg}$). The solid white line indicates the gate voltages at which the superconducting quantum interference is investigated in the main text.
}
	\label{SQI:figS2}
\end{figure}

\section{Hall bar measurement}
A Hall bar device with length 80 \textmu m and width 20 \textmu m is fabricated on the same wafer. The mobility and density as a function of top and bottom gate are shown in Fig.\,S3. The following parameters are extracted and calculated at $V_{tg}$ = 0 V and $V_{bg}$ = 0 V, using the effective mass of InAs $m^* = 0.04$ $m_e$ \cite{Qu_2015} and Lande g-factor 11.5 \cite{Mu_2016}.
\begin{align*}
&n =\, 1.1 \cdot 10^{16}\, \mathrm{ m^{-2}} \\
&\mu =\, 1.6 \cdot 10^5\, \mathrm{ cm^2 / (V s)} \\
&v_F =\, \hbar \sqrt{2 \pi n}/m^* = 7.6 \cdot 10^5\, \mathrm{ m/s} \\
&\lambda_{MFP} =\, \frac{\hbar \mu}{e} \sqrt{2 \pi n} =  2.8\, \mathrm{\mu m} \\
&E_{thouless} =\, \frac{h v_F}{L} = 6.3\, \mathrm{meV} \\
&E_{zeeman} =\, g \mu_B B \overset{\mathrm{10 mT}}= 0.0066\, \mathrm{meV} \\
\end{align*}

\begin{figure}[!ht] 
	\centering
	\includegraphics[width=7 in]{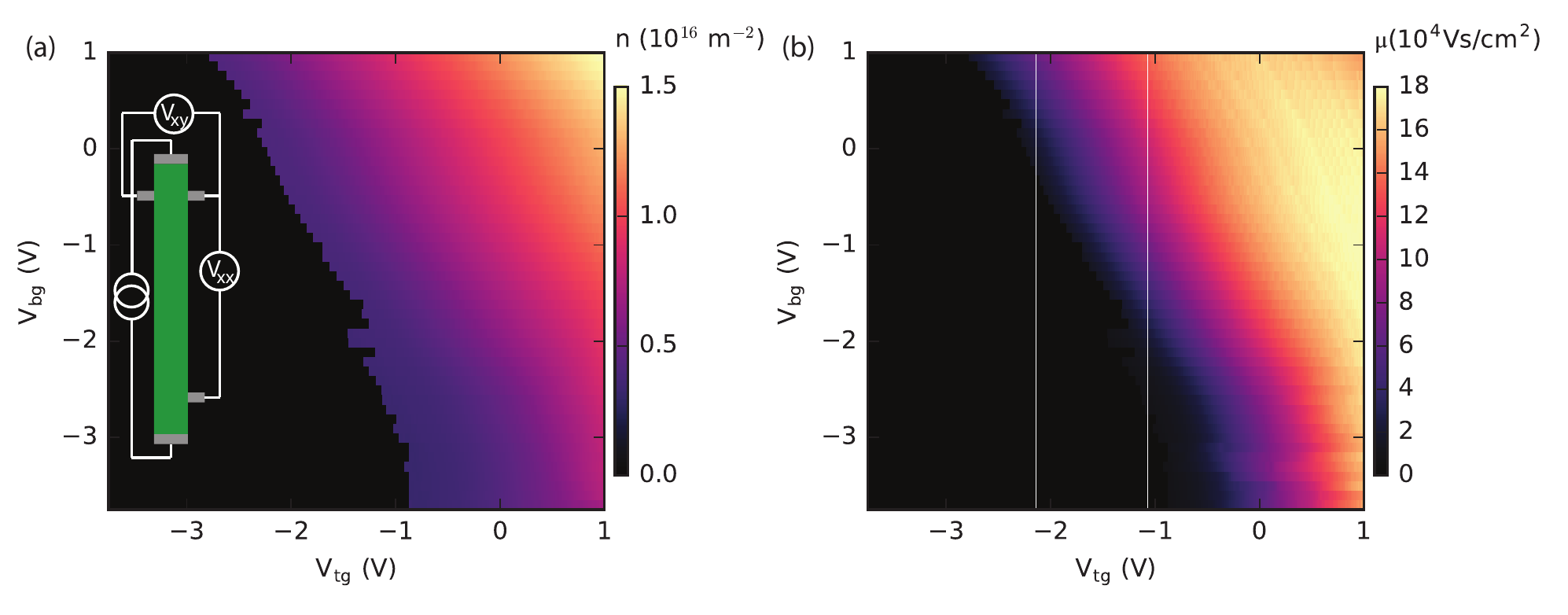}
	\caption{(a) The density of the Hall bar device (inset) is plotted as a function of top gate and bottom gate voltage. Each line is cut off at the point where the density became too low to measure. In (b) the mobility is plotted for the same gate voltage ranges as in (a).
}
	\label{SQI:figS3}
\end{figure}

\section{Flux focusing and vortices}
The contact geometry causes a part of the flux to be focussed in the junction due to the Meissner effect. The NbTiN contacts have a width of 4 \textmu m and a length of 1 \textmu m. If there are no vortices in the NbTiN (type II superconductor), the maximum ammount of flux will be threaded to the junction. If there are vortices, a flux quantum per vortex penetrates the superconductor and will therefore not be focused through the junction. Experiments on strip geometry Nb films show that the critictal field at which vortices appear in te strip is given by $B_m = \Phi_0 / W^2$ \cite{Stan_2004_SM}, in our sytem equal to 2 mT. We assume that the vortices settle at the beginning of a measurement, since the magnetic field is swept relatively fast to the starting point, whereafter it is changed in small steps before a current bias trace is measured. Then different sweep ranges will have different periodicities, since the number of vortices alter the flux focusing. More specifically, smaller sweep ranges will have smaller periodicity, because there are less vortices and therefore more flux focusing. This is what we see if we compare the flux periodicities in Fig.\,2(b) and Fig.\,3(a) in the main text. To reassure this observation we show a wider range magnetic field scan of the SQUID regime in Fig.\,S4. The observed periodicity is used to extract an effective length of 1.2 \textmu m, which is the same as for the Fraunhofer pattern in Fig.\,2(b) and even-odd pattern in Fig.\,4(a).
\begin{figure}[!ht] 
	\centering
	\includegraphics[width=6 in]{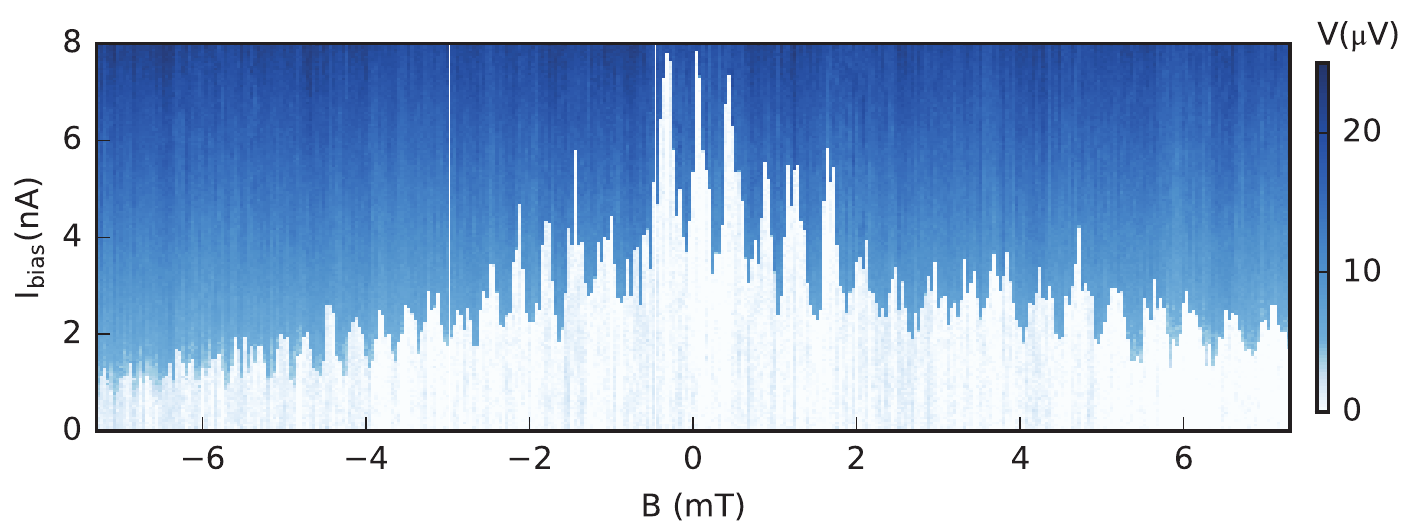}
	\caption{SQI measurement at $V_{tg}$ = -7 V and $V_{bg}$ = -1.15 V. The observed SQUID pattern has a period of 0.4 mT, from which an effective length of 1.2 \textmu m is estimated.
}
	\label{SQI:figS4}
\end{figure}

\newpage
\section{Fraunhofer to SQUID in two other devices}
\begin{figure}[h!] 
	\centering
	\includegraphics[width=7 in]{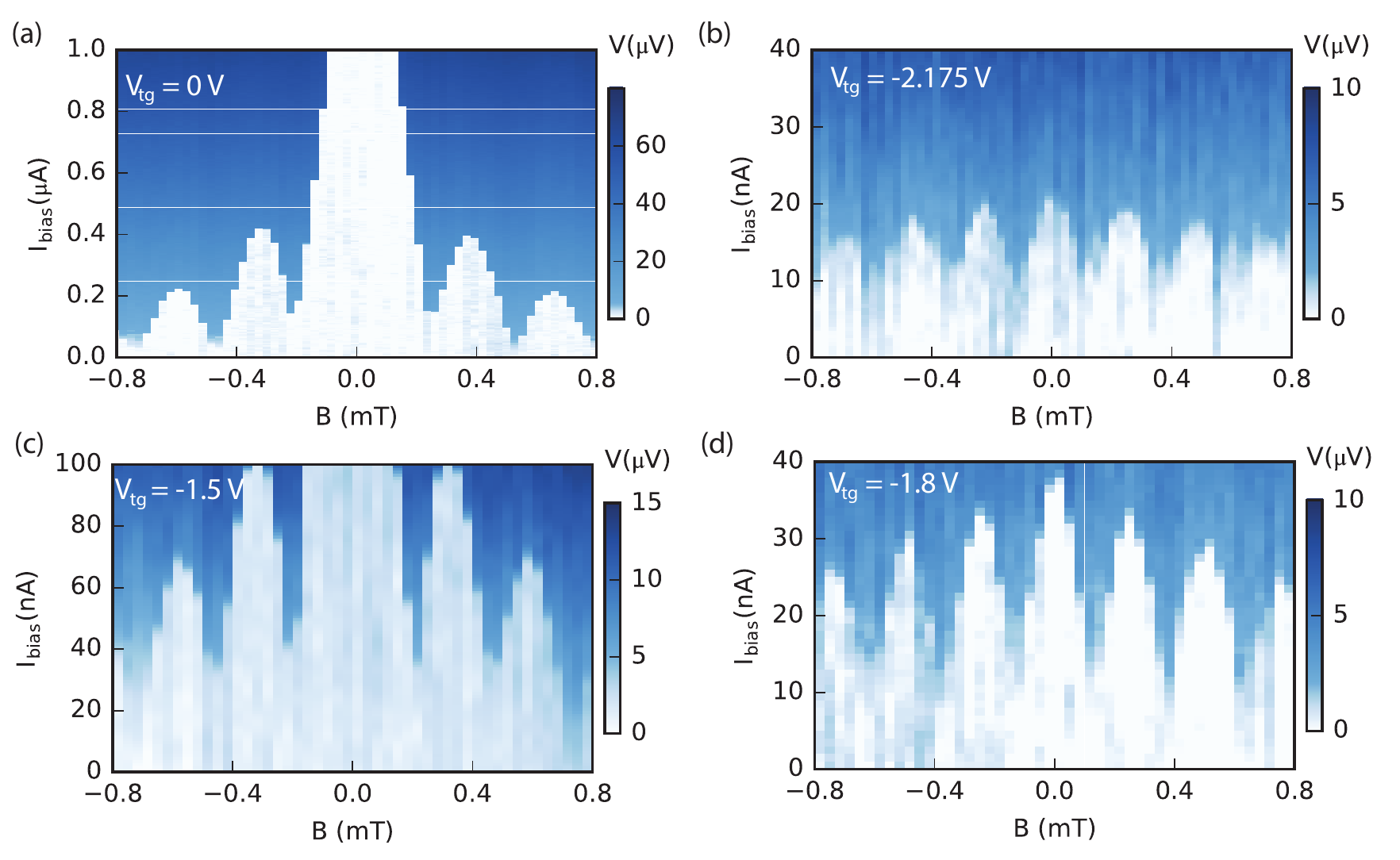}
	\caption{SQI data at the indicated top gate voltages is shown of two lithographically similar devices ($L$ = 500 nm, $W$ = 4 \textmu m). The Fraunhofer SQI patterns are shown in (a) and (c), and the SQUID patterns in (b) and (d) for device 2 and 3, respectively. For all figures $V_{bg}$ = 0 V. Clearly in both devices the SQI pattern turns from Fraunhofer to SQUID by decreasing the gate voltage.
}
	\label{SQI:figS5}
\end{figure}
\newpage

\section{Simulation contact edge}
\begin{figure}[h] 
	\centering
	\includegraphics[width=5.2 in]{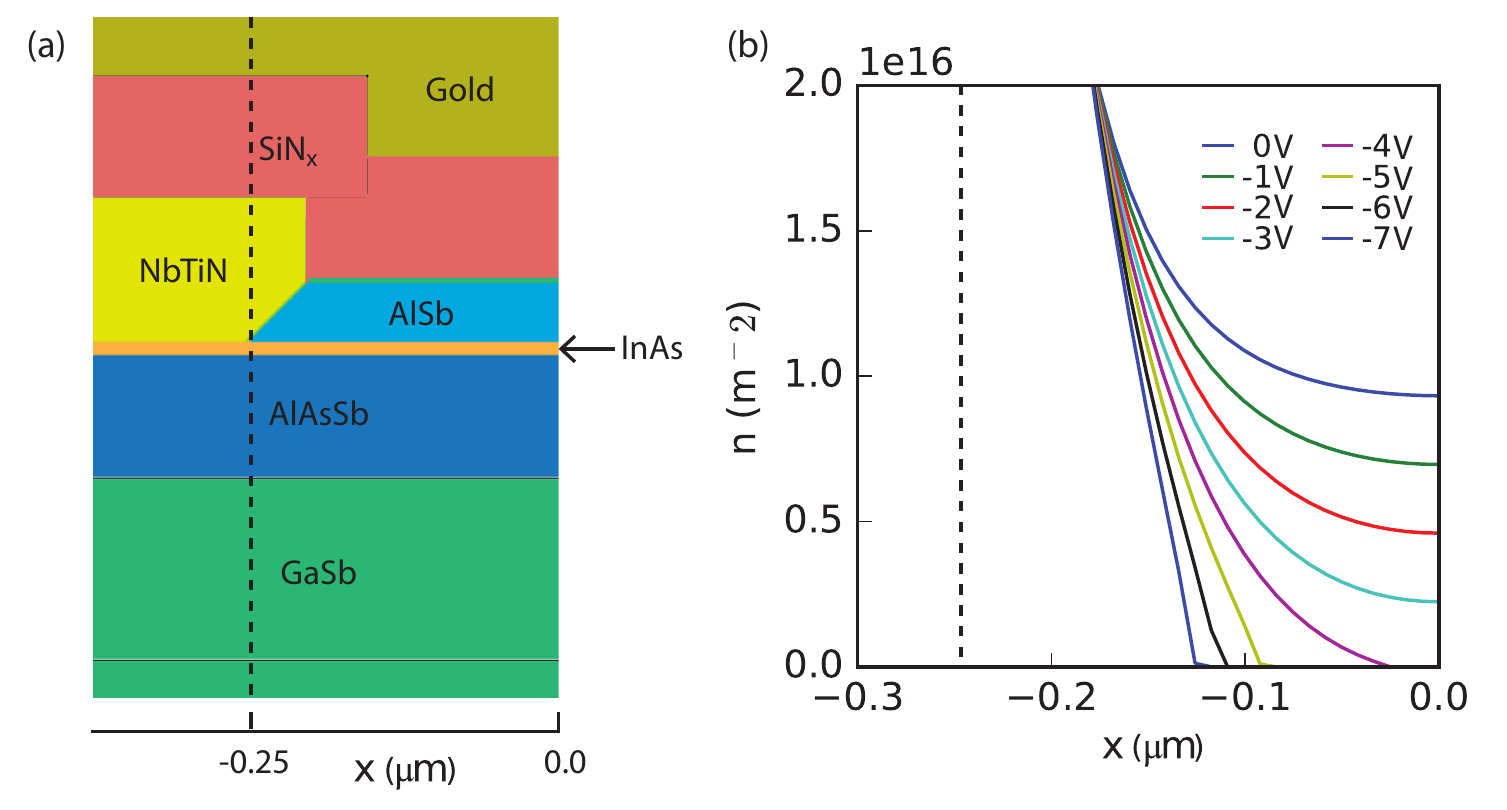}
	\caption{ Electrostatic simulations are performed using a finite element Poisson solver. (a) The device geometry as used in the simulation. The SiN$_x$ dielectric is 100 nm thick and the NbTiN 120 nm, all other thicknesses can be found in Fig\,S1. (b) The electron density profile is changing as the top and bottom gate voltages are swept. The top gate voltage is indicated in the legend. In the lower gate voltage traces the electrons in the bulk are depleted while there is still a large electron density in the vicinity of the contacts. For claryfication, the x-axis value where the NbTiN contact is on top of the InAs is indicated by the dashed line in both (a) and (b). The top gate is screened close to the contacts due to the triangular shape of the top AlSb barrier. This shape is caused by isotropic wet etching of the mesa.
}
	\label{SQI:figS6}
\end{figure}
\newpage

\section{Extended Dynes-Fulton analysis}
\begin{figure}[H] 
	\centering
	\includegraphics[width=6 in]{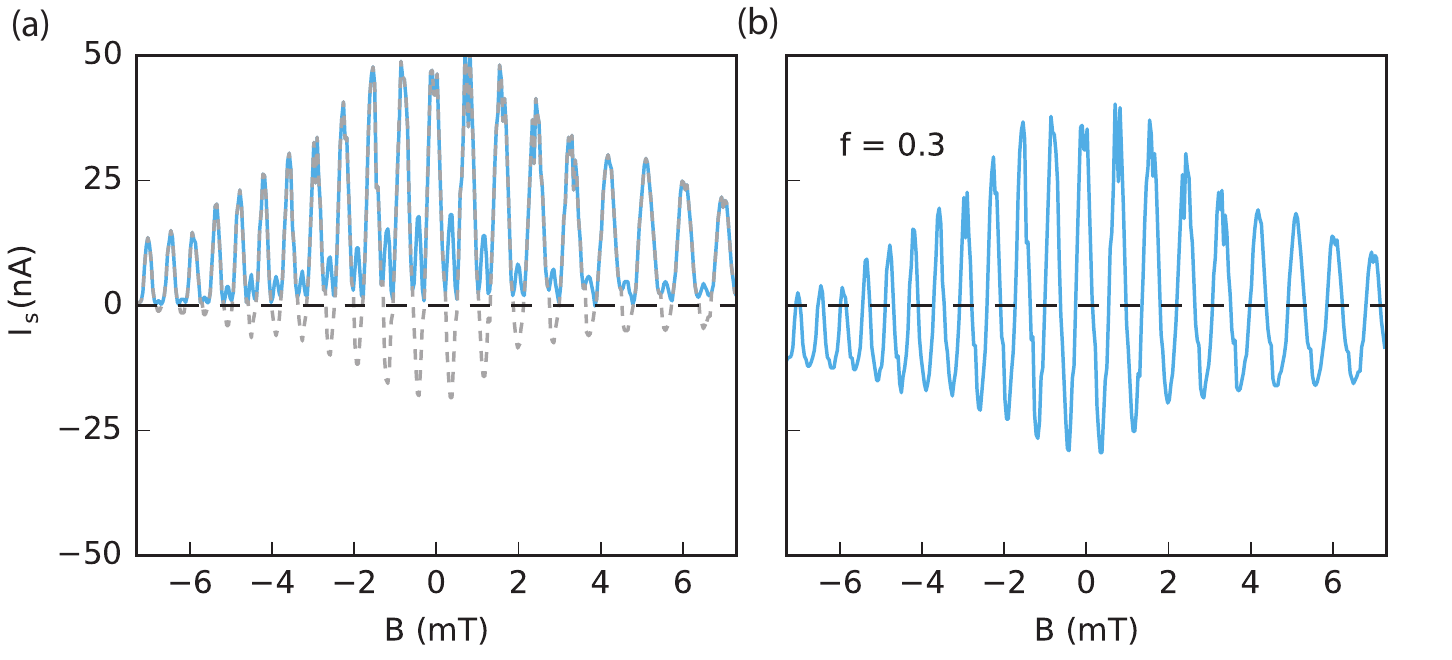}
	\caption{(a) The blue line is the switching current as a function of magnetic field as extracted from the measurement of Fig.\,4(a). The grey dashed line depicts the pattern after flipping every other node as is usually done in the Dynes-Fulton analysis \cite{Dynes_1971_SM}. Taking into account crossed Andreev reflection and the resulting inteference effects, we expect the SQI pattern to have the form of equation (1) in the main text, the absolute value of a vertically offsetted cosine. (b) shows the same pattern as (a) with an offset of 11 nA subtracted, equal to a factor of 0.3 times the switching current at zero magnetic field.
}
	\label{SQI:figS7}
\end{figure}
\newpage

\section{Tight binding model}

\begin{figure}[H]
  \centering
  \includegraphics[width=0.8\linewidth]{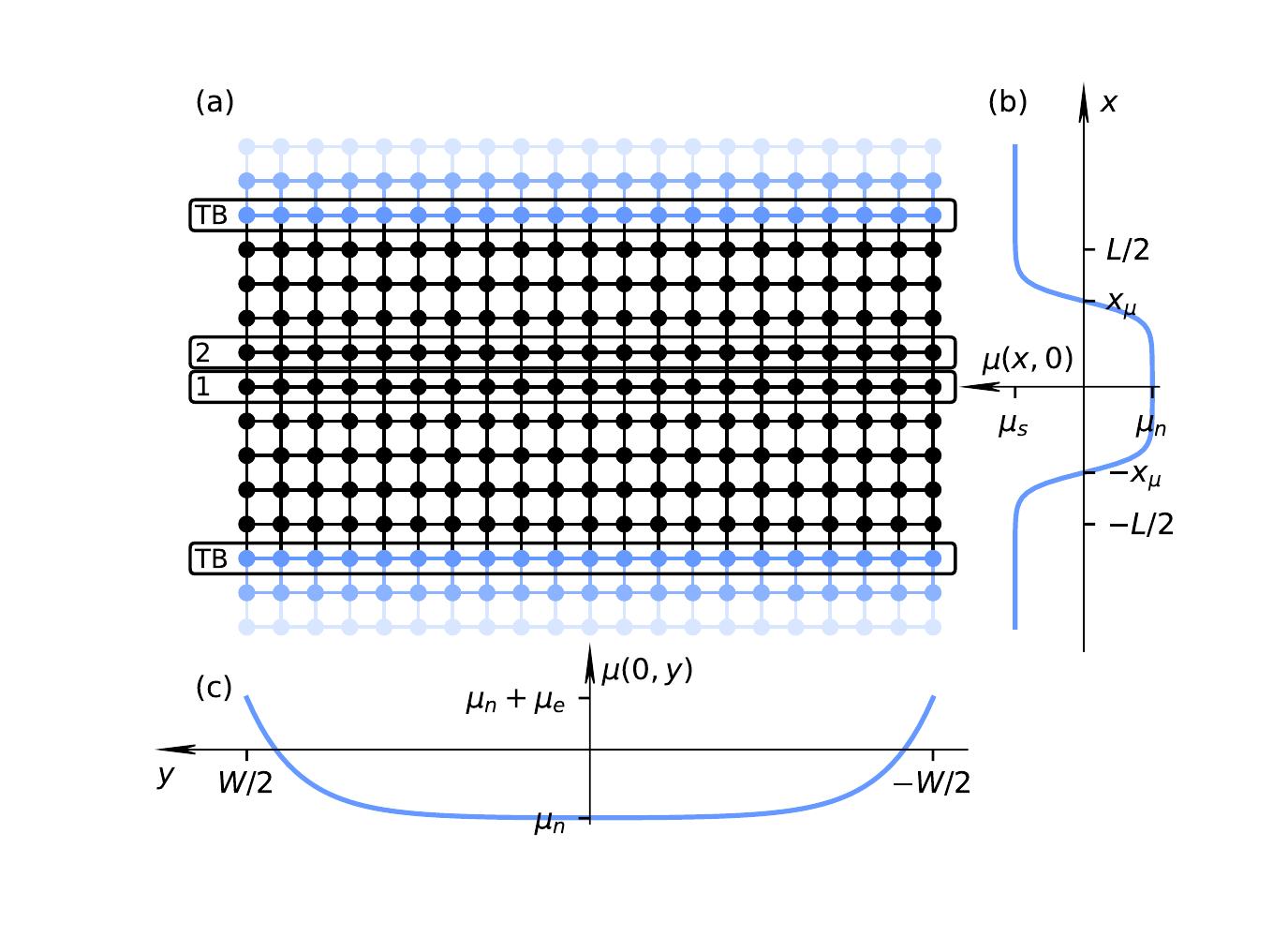}
  \caption{
    (a)~Schematical representation of a tight-binding model.
    Bogolyubov-de Gennes Hamiltonian is discretized on a square lattice.
    Superconducting sites of the system have a blue color, normal -- black.
    A tunnel barrier is created, using one row of sites with decreased chemical potential (marked TB on the scheme).
    The current was calculated from Green's function of sites, marked 1 and 2 on the scheme (see the detailed explanation
    below).
    (b)~Chemical potential profile for $x=0$.
    Offset between location of chemical potential step and superconducting region together with the tunnel barrier leads
    to formation of scattering channel between edges.
    (c)~Chemical potential profile for $y=0$.
    Band bending is represented with an increased chemical potential at the edges, leading to edge conductivity in a
    doped regime.
  }
  \label{fig:sup_tb_scheme}
\end{figure}

We have taken the following Hamiltonian for tight binding simulations:
\begin{equation}
  H = \left(\frac{\hbar^2 (k_x^2 + k_y^2)}{2 m_\text{eff}} - \mu(x, y)\right) \tau_z +
    \alpha (k_x \sigma_y - k_y \sigma_x) \tau_z + g \mu_\text{B} B(x) \sigma_z + \Delta(x) \tau_x,
\end{equation}
where $\sigma$ Pauli matrices correspond to the spin degree of freedom, and $\tau$ -- to the electron and hole one.
It is discretized on a square lattice with lattice constant $a = 2 \text{ nm}$.
The normal part of a SNS junction is represented as a rectangle $-L/2 \le x \le L/2$ and $-W/2 \le y \le W/2$,
the superconducting parts -- as translationally invariant in $x$ direction stripes with $-W/2 \le y \le W/2$.
Proximity-induced pairing potential $\Delta(x)$ is zero in a normal part and constant in a superconducting
part of the system, with a step-like transition.
The magnetic field is assumed to be fully screened by the superconductors. Its impact is included as Zeeman term and via Peierls substitution.

At first realistic values of $\alpha =  5 \cdot 10^{-12}\,\text{eV} \cdot \text{m}$ and $g = 11.5$ for the Rashba and Zeeman term were used to verify that they do not play an important role in this parameter regime. After we were sure that Zeeman and Rashba terms can be neglected, we have put $\alpha=0$ and $g=0$  for the sake of numerical performance.
This allowed to decouple spins and decrease the dimensionality of the Hamiltonian twice, since both decoupled subblocks
contribute equally to the current.

Chemical potential $\mu(x, y)$ is selected to capture primary features of the device: band bending near the edges and
screening near the NS boundaries top gate.
It has the following form:
\begin{align}
  \mu(x, y) = {}&\frac{\mu_\text{norm} + \delta\mu_\text{edge}(y)}{2} \left(\tanh\frac{x + x_\mu}{\lambda_\mu} -
    \tanh\frac{x - x_\mu}{\lambda_\mu}\right) + \nonumber\\
  {} &\frac{\mu_\text{sc}}{2} \left(2 - \tanh\frac{x + x_\mu}{\lambda_\mu} + \tanh\frac{x - x_\mu}{\lambda_\mu}\right),
\end{align}
where
\begin{equation}
  \delta\mu_\text{edge}(y) = 2 \mu_\text{e} e^{-W/2\lambda_e} \cosh \frac{y}{\lambda_e}
\end{equation}
is the term, that introduces band bending near the edges of a normal part.
$\mu_\text{norm}$ and $\mu_\text{sc}$ are chemical potentials in gated area (primarily normal part) and area screened
by the superconducting contacts.
If normal part is governed to the insulating state with negative $\mu_\text{norm}$, the offset between $L/2$ and $x_\mu$ leads
to formation of a conducting channel on the NS boundaries of the junction, with a width:
\begin{equation}
  W_\text{ns} = L/2 - x_\mu.
\end{equation}
The tunnel barrier on the NS interface was represented as a single row of sites with a chemical potential reduced by $\Delta\mu_\text{TB}$.

The finite-temperature critical current of the SNS junction was calculated by maximizing the current-phase dependency, similarly
to the approach, used in~\cite{furusaki_dc_1994}.
The Green's function was numerically calculated for several Matsubara frequencies on two neighbouring rows of the sites in
the normal part of the junction, then the current was obtained by the summation:
\begin{equation}
  I = \frac{2 e k_B T}{\hbar} \sum_{n=0}^{N_\text{max}} \left( \Im \tr H_{21} G_{12} (i \omega_n) -
  \Im \tr H_{12} G_{21} (i \omega_n) \right).
\end{equation}
Here $H_{21}$ and $G_{21}$ denote hopping matrix and Green's function subblock from cells of row~1 to row~2, indicated
on Fig.~\ref{fig:sup_tb_scheme} (all the hoppings, that form a cut through the system), and vice versa.
$\omega_n = (2 n + 1) \pi k_B T$ is the $n$-th Matsubara frequency.
Value $N_\text{max}$ was obtained dynamically, based on the estimated convergence rate.
The Green's functions were calculated, using package Kwant~\cite{kwant_paper}.

The numerical values of parameters, used for simulations, are presented in Table~\ref{tab:num_paramters}.
A lattice constant of $a = 2\text{ nm}$ was selected small enough to capture characteristic length scales of an edge and NS
boundary current channels.\\

\begin{table}[h]
  \centering
  \begin{tabular}{|c|c|c|c|c|c|c|c|c|c|c|}
    \hline\hline
    $W\,[\text{nm}]$ & $L\,[\text{nm}]$ & $\lambda_\text{e}\,[\text{nm}]$ & $\lambda_\mu\,[\text{nm}]$ & $x_\mu\,[\text{nm}]$ &
    $m_\text{eff}/m_\text{e}$ &  $\Delta \, [\text{eV}]$ & $\mu_\text{sc} \, [\text{eV}]$ & $\delta\mu_\text{e}
    \, [\text{eV}]$ \\ \hline
    $400$ & $200$ & $28$ & $1$ & $0 \div 50$ & $0.04$ & $4 \cdot 10^{-4}$ & $0.2$ & $0.15$ \\
    \hline\hline
  \end{tabular}
  \caption{
    Numerical parameters, used for tight-binding simulations.
  }
  \label{tab:num_paramters}
\end{table}
\newpage

\section{Tunnel barrier dependence in tight binding model}
\begin{figure}[ht] 
	\centering
  \includegraphics[width=5.2 in]{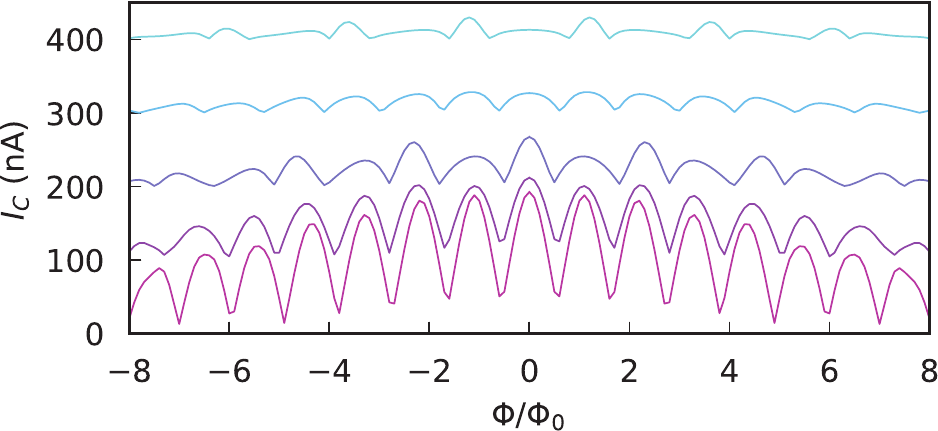}
	\caption{Tight binding calculation of the superconducting quantum interference as a function of tunnel barrier strength at the contact. Increasing the tunnel barrier height leads to enhanced normal reflection with respect to Andreev reflection. The electrons or holes then have a higher chance of traversing along the contact before Andreev reflecting. Forming a crossed Andreev states requires the charge carriers to traverse around the junction fully. Therefore enhanced normal reflection is benefecial for forming these states and the resulting even-odd SQI pattern. Here we plot the SQI patterns for a tunnel barrier strength ranging from 0.6 eV to 1.40 eV (bottom to top) in 0.2 eV steps.
}
	\label{SQI:figS8}
\end{figure}
\newpage

\section{$W_{ns}$ dependence in tight binding model}
\begin{figure}[H] 
	\centering
  \includegraphics[width=5.2 in]{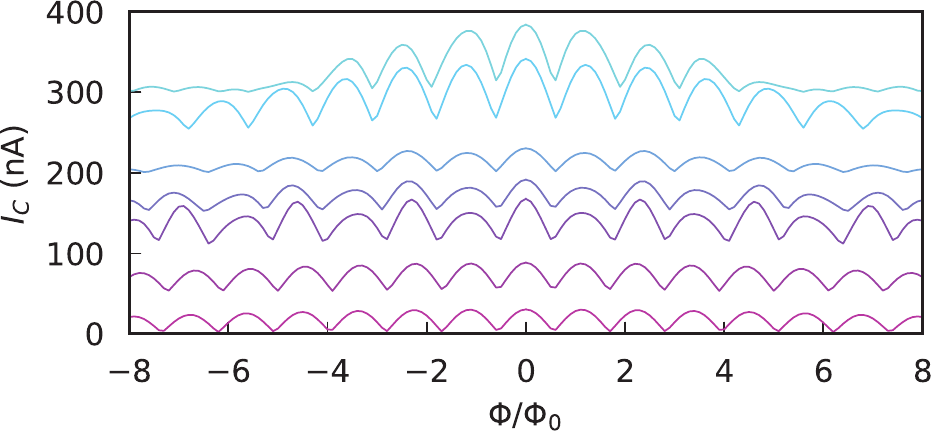}
	\caption{Tight binding calculation of the superconducting quantum interference as a function of width of the channel along the edge. As a sanity check: if the width is 2 nm (bottom trace), we do not see even-odd effect. Increasing the width (in 8 nm steps up to 50 nm), increases the number of channels along the contact and the coherence length, up to the point that the 1D channel become 2D and the even-odd effect reduces again.
}
	\label{SQI:figS9}
\end{figure}

\end{document}